# Generation and subwavelength focusing of longitudinal magnetic fields in a metallized fiber tip

Daniel Ploss,[1,*] Arian Kriesch,[1] Hannes Pfeifer,[1] Peter Banzer,[1,2] and Ulf Peschel[1]

[1]*Institute of Optics, Information and Photonics, Erlangen Graduate School in Advanced Optical Technologies, EAM, Friedrich-Alexander-University Erlangen-Nuremberg, Haberstr. 9a / Staudtstr. 7 B2, 91058 Erlangen, Germany*
[2]*Max Planck Institute for the Science of Light, Guenther-Scharowsky-Str. 1 Bldg. 24, 91058 Erlangen, Germany*
[*]*daniel.ploss@mpl.mpg.de*

**Abstract:** We demonstrate experimentally and numerically that in fiber tips as they are used in NSOMs azimuthally polarized electrical fields ($|\mathbf{E}_{azi}|^2 / |\mathbf{E}_{tot}|^2 \approx 55\% \pm 5\%$ for $\lambda_0 = 1550$ nm), respectively subwavelength confined (FWHM $\approx 450$ nm $\approx \lambda_0/3.5$) magnetic fields, are generated for a certain tip aperture diameter ($d = 1.4$ μm). We attribute the generation of this field distribution in metal-coated fiber tips to symmetry breaking in the bend and subsequent plasmonic mode filtering in the truncated conical taper.

## 1. Introduction

Near-field scanning optical microscopy (NSOM) aims for the experimental investigation of near-field distributions around nanostructures [1–4] mainly because resolving subwavelength features via the far field is not possible due to the Abbe diffraction limit [5,6]. However, the vectorial nature of the electromagnetic field with its distribution of electric and magnetic components exhibiting subwavelength features contains far more information than just the field amplitude [7]. Every image recorded with an NSOM is inevitably influenced by the vectorial field-distribution induced in the probe and by that of the sample. Exact knowledge of the optical properties of the tip in terms of its near-field polarization distribution is therefore essential for understanding the field distribution around the sample [8–10]. Moreover, there have been first demonstrations showing that the aperture of an NSOM probe can be engineered to selectively detect certain field components, e.g. the magnetic moment of the electromagnetic field of a sample that was scanned [11–13] and in-plane electric fields were selectively detected with deep-subwavelength resolution [14]. It has also been shown numerically that corrugated metal-coated silica tips [15,16] can act as efficient plasmonic concentrators and can in principle focus externally generated azimuthally polarized light down to a subwavelength spot (FWHM ≥ $\lambda_0/19$) of a longitudinally polarized magnetic field [17]. However, as the closest comparable experiment to date, the current benchmark is set for sensing such field distributions with a sophisticated fabrication technique by Kuipers et al. [11,18]. In this paper we demonstrate experimentally and theoretically that by just cutting a NSOM tip at the right diameter we generate and concentrate a magnetic field to an extremely small focus (FWHM ≈$\lambda_0/3.5$).

As the field propagation in linear optics is time-reversible there is no fundamental difference between the case where light is collected from the sample via the tip (collection mode) and the situation where the sample is locally excited by light emitted from the tip (illumination mode). Hence, by analyzing the field emitted by a typical fiber aperture NSOM tip [19–21] we also gain information about the sensitivity of the tip with respect to certain vectorial components detected in collection mode.

Here we experimentally show that a linearly polarized mode ($\lambda_0$ = 1550 nm) in an optical fiber is converted into partially azimuthally polarized light ($|\mathbf{E}_{azi}|^2 / |\mathbf{E}_{tot}|^2$ ≈55% ± 5%) by passing through a tapered, bent and metal-coated aperture NSOM tip (tip aperture diameter $d$ = 1.4 µm). At the same time the highest transmission through the NSOM tip is obtained when a maximum portion of azimuthal polarization is measured at the tip apex. For comparison we perform a numerical analysis and discuss the field evolution in terms of symmetry considerations. It turns out that symmetry breaking introduced by the fiber bend plays a key role in the conversion process. Light leaving the fiber tip at its very end is confined to a tiny spot that is smaller than the wavelength and can be positioned with the resolution of a few nanometers. This makes the presented optical setup an easy to reproduce source for azimuthally polarized coherent light. Tightly focused azimuthally polarized free-space beams have already found a variety of applications in nanooptics and plasmonics [22–24]. With the presented approach the size of these experimental setups can potentially be dramatically reduced and robustness can be gained with little technical effort. By fiber-optical integration of all relevant parts this system is easily kept stable over long timescales.

## 2. Experimental setup and measurement results

The investigated NSOM tips are produced by drawing, bending and metallizing the end of an optical fiber and are subsequently mounted to a piezo driven tuning fork. Therefore each tip is intrinsically connected with a 2 m long piece of single mode optical fiber for $\lambda_0$ = 1550 nm (SMF 28). The raw tips are manufactured by Nanonics Inc. and supplied for a Nanonics MultiView MV 4000 NSOM system. To launch light into the NSOM tip we prepared a homogeneously polarized Gaussian laser beam ($\lambda_0$ = 1550 nm) with an accurately controllable polarization state using a linear polarizer followed by a half- and a quarter-wave-plate. This beam was coupled into the single mode fiber. As polarization coupling might occur along the fiber we have no direct information about the actual field distribution at the entrance to the tip, still it must be a superposition of the two linearly polarized modes of the fiber, the relative phase and amplitude of which we can vary in a controlled way by adjusting the input polarization.

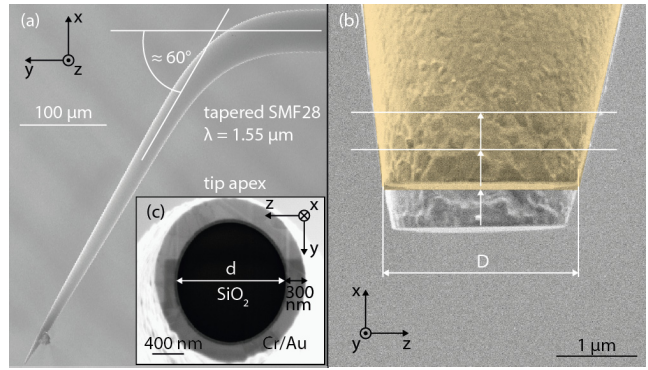

Fig. 1. (a) SEM of a tapered, bent and metal-coated (Cr/Au) NSOM tip. At the beginning of the bend the outer apex diameter $D$ is ≈50 μm, which tapers to $D$ ≈34 μm after a ≈60° bend. Subsequently a truncated cone with a length of ≈360 μm follows until the tip apex. (b) Colorized SEM of the stepwise FIB cut-back to increase the lateral outer apex diameter $D$. (c) Polished tip apex for a tip aperture diameter of $d$ = 1.4 μm.

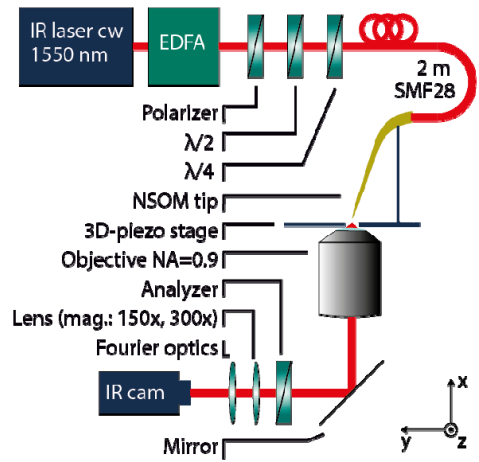

Fig. 2. The experimental setup for imaging the NSOM tip polarization-controlled emission in real and Fourier space. The light from the 1.5 mW fiber-coupled IR laser is amplified to a total power between 9 mW and 11 mW. The power is sent into a free-space setup and is directed through a polarizer, a half- and a quarter-wave-plate before it is fiber-coupled into the 2 m long single mode fiber (SMF 28) with the NSOM tip at the end.

The fiber end is tapered down from its original fiber diameter of 125 μm to 50 μm, then bent by about 60° and again tapered to an aperture diameter $d$ = 432 nm - 2.2 μm, as determined from SEM images (Fig. 1). The whole fiber end is coated with metal (Au-Cr) except a small opening at the tip apex. This tip apex can be positioned in x, y and z direction with a resolution of 5 nm using a stabilized piezo scanning stage (PI Instruments, P-527.3CD) included in the custom made optical setup (Fig. 2) that follows the design published by Banzer et al. [23]. The tip aperture was imaged onto an InGaAs CCD camera (Xenics XS, 320 x 256 pixels) with 150x/300x magnification using a microscope objective (NA = 0.9). If required, additional optical components were inserted into the imaging beam path to investigate the angular emission spectrum of the tip by recording the Fourier plane. The polarization state of the emission was investigated by rotating a polarization filter (analyzer) in front of the camera or imaging optics.

The elliptic polarization state that was coupled into the fiber was iteratively adjusted for maximum transmission through the NSOM tip, while observing the total output power. After characterizing the transmission properties of the tip we stepwise cut-back the tip with a focused Ga ion beam system (FIB) thus successively increasing the diameter of the tip aperture (Figs. 1(b) and 1(c)). The potentials of FIB processing of NSOM tip apexes were recently demonstrated [25–27]. We always ensured that the tips were clean-cut in a plane parallel to the NSOM tip scanning plane and polished with a low-current focused ion beam [14]. The emitting tip apex plane was therefore always tilted by ~30° with respect to the direction of the light propagation in the tapered waveguide. As our simulations have revealed this tilt reduces the portion of the azimuthally polarized components at the tip aperture slightly, which will be explained in section 3. The whole cut-back procedure was carefully controlled and respective diameters were determined using a charge-effect-free parallax-free high magnification scanning electron microscope (SEM) (Fig. 1(b)). For every tip diameter all steps for optical transmission characterization were repeated including the optimization of the polarization state of the launched light for maximum transmission. All experiments were performed on a set of four different NSOM tips of the same type, all of which showed consistent results.

We started at a tip aperture diameter of $d = 1.0$ μm and opened the tip successively to $d = 1.4$ μm, $d = 1.8$ μm and $d = 2.2$ μm. For $d = 1.0$ μm we observed the emission of linearly polarized light without any significant additional higher order modes being present (Figs. 3(a) and 3(d)). The dominance of linear polarization for small tips is consistent with published experimental results obtained for smaller tip aperture diameters of $d = 100$ nm [28] and for published theoretically results for $d = 30$ nm with linear s and p polarized light [29].

The emitted polarization changed fundamentally for a tip aperture diameter of $d = 1.4$ μm (Figs. 3(b) and 3(e)). The polarization-resolved far-field image of the tip aperture clearly shows a partially azimuthally polarized field distribution (Figs. 3(g) and 3(h)). This seems to result from a superposition of two modes of the circular metallized fiber tip: a quasi-azimuthally and a quasi-linearly polarized mode (polarized in z-direction, see Fig. 3(e)). Due to interference between both modes an asymmetry in the intensity in y-direction is induced (Figs. 3(b) and 3(e)). For larger diameters, starting with $d = 1.8$ μm, an azimuthally mode contribution remains detectable (Figs. 3(c) and 3(f)). However, it becomes increasingly obfuscated by an increasing background of successively evolving higher order modes that are only supported by tip aperture diameters larger than $d = 1.4$ μm. Hence, for a wavelength of 1.55 μm the NSOM tip aperture diameter of $d = 1.4$ μm is the natural choice to achieve a dominant azimuthal polarization.

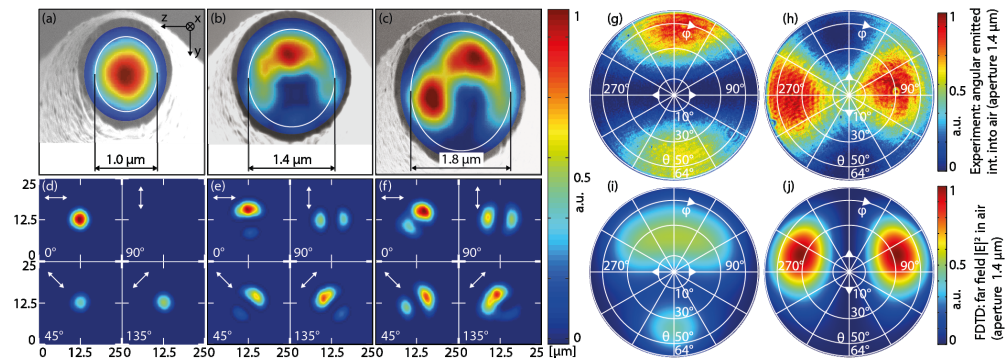

Fig. 3. Measured field intensity at the tip apex overlaid with the respective SEM picture for (a) $d = 1.0$ μm, (b) $d = 1.4$ μm and (c) $d = 1.8$ μm tip aperture diameter. The white ellipses mark the tip aperture border between the outer metal and the inner silica cladding. (d) – (f) Field intensities at the tip apex imaged through a polarizer of varying orientations for the tip aperture diameters displayed above. (d) For $d = 1.0$ μm the measurement shows linear polarization in z-direction. (e) For $d = 1.4$ μm partially azimuthal polarization is visible. (f) Higher order modes are visible for $d = 1.8$ μm. (g) – (h) Directionality of the emitted farfield for $d = 1.4$ μm for horizontal and vertical polarization compared with (i) – (j) numerical simulations.

The experimental observation of azimuthal polarization at the tip apex of NSOM tips has not been reported before. In addition, the transformation of a linearly polarized fiber mode into an azimuthal polarization state is a-priori unexpected, as this kind of conversion inevitably requires the breaking of cylindrical symmetry. Different to sections of the tapered single mode optical fiber where such polarization conversion is forbidden, cylindrical symmetry is broken in the fiber bend, which therefore plays a crucial role in the conversion process. The whole tip geometry lacks cylindrical symmetry, but is mirror symmetric with respect to the xy-plane (Fig. 1(a)). The experimentally determined distribution of the electric field at the tip apex is antisymmetric, hence its mirror image at the xy-plane equals the negative of the original field. Therefore also the input field above the bend must obey this symmetry condition. Hence, in the case of maximum transmission only an asymmetric fiber mode with linear polarization normal to the symmetry plane (xy) is present at the entrance of the bend.

## 3. FDTD simulation

Taking these constraints into account, we modeled the whole system beginning with the bend at a fiber diameter of 50 μm (Fig. 4). Although this is less than half of the diameter of the original fiber still two linearly polarized modes are guided in the fiber core and can be found numerically with a mode solver. A respective z-polarized mode served as the input field of further propagation through the metal-coated bend and the subsequent taper (Figs. 4(I)–4(III)) down to the wavelength-scale emitting aperture. Here the change from a multimode regime of the propagating light inside the tapered fiber to the cutoff close to the tip apex is clearly visible (Fig. 4(b)).

Although we made use of the mirror symmetry of the configuration the achievable resolution of 3D finite difference time domain (FDTD solutions by Lumerical Solutions Inc) simulations was limited by the macroscopic dimensions of the structure, leading to a large number of numerical Yee-nodes. Reasonable accuracy combined with still tolerable computation times were achieved with a rectangular mesh of 100 nm spacing ($\lambda_0 = 1550$ nm). The silica core of the tip was modeled as lossless fused silica ($n = 1.444$) [30]. As the tips were fabricated by drawing a single mode fiber down to the required diameter, we assumed a region of slightly increased refractive index of $\Delta n = 5 \cdot 10^{-3}$ in the center of the structure [31]. The diameter of that core was derived from a standard fiber (core diameter: 8.2 μm, cladding diameter: 125.0 μm ± 0.7 μm) [31] by assuming volume conservation during drawing. The metal cladding was modeled as a 300 nm thick layer of Au ($\varepsilon = -116.3 + 11.6i$) [32]. The Cr contact layer was neglected because of its small thickness of only 3 nm.

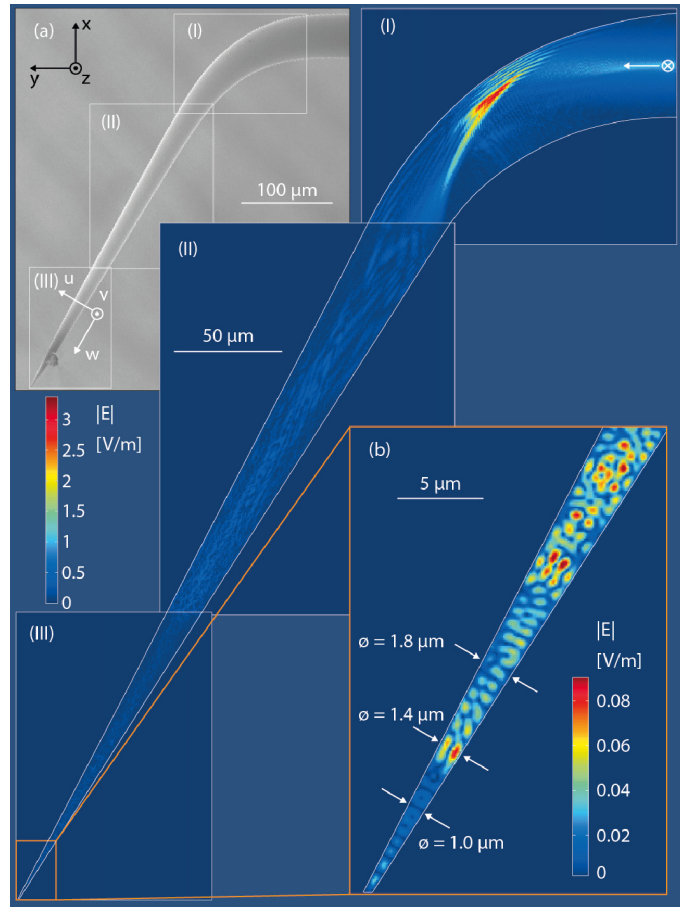

Fig. 4. Field evolution in a bent and metal-coated NSOM tip. (a) SEM, which formed the basis of the 3D FDTD. (I-III) The simulation was split into 3 domains. (I) Light of a linearly polarized mode is reflected on the concave bend wall into the tapered tip (II-III). (b) As the tip diameter decreases higher order modes become successively extinct. Close to the tip apex only one azimuthally and one linearly polarized mode are supported.

At the beginning of the bend, the fiber is already tapered down to less than half of its original diameter and therefore light is only weakly guided in the fiber core. Hence, the field in the bend propagates straight along the y-axis, thus leaving the fiber core that has no significant effect on any further field propagation. After it has left the fiber core, the light is reflected by the metal-coated inner surface of the tip that forms a concave, asymmetric mirror. Azimuthal components of the field are generated in the course of this first reflection process (Fig. 4(I)). Higher order modes and in particular azimuthal ones are generated during this reflection and propagate down the taper. As the taper itself has cylindrical symmetry, no further mode conversion occurs and all modes propagate down the taper until a certain mode dependent diameter is reached. Higher order modes experience this mode cutoff much earlier and finally only a linearly polarized and the lowest order azimuthally polarized mode reach tip aperture diameters below $d = 1.5$ μm. During propagation, radial field components penetrate deeper into the metal, thus experience stronger absorptive losses. Therefore the azimuthal mode is less affected by losses and its relative content increases along propagation until the mode reaches its cutoff.

In order to quantify the amount of power guided in the azimuthally polarized lowest order mode we determined the strength of the azimuthal field component as

$$E_\phi(r) = \frac{1}{2\pi} \int_0^{2\pi} d\phi \, E(r,\phi) \, e_\phi \tag{1}$$

and calculated the respective power ratio as

$$\gamma = \frac{\int_0^\infty dr \, 2\pi r \, |E_\phi(r)|^2}{\int_{-\infty}^\infty dx \int_{-\infty}^\infty dy \, |E(x,y,z)|^2} \tag{2}$$

(Fig. 5(a)) using monitors in the u,v,w-coordiante system (Fig. 4(a)) of the taper. In agreement with our experimental results, a maximum of roughly 58% of azimuthal power content appears for a tip aperture diameter of $d = 1.4$ μm. The remaining 42% of power are contained in the linear polarized mode. As mentioned in the beginning, the real NSOM tip scanning plane is tilted by ~30° and lies in the y,z-plane (see Fig. 4(a)). Here, a maximum of roughly 54% azimuthal power content appears for that tip aperture diameter, which is still in good agreement with the experimental results. For even smaller diameters the content of azimuthally polarized light drops rapidly as the respective mode approaches its cutoff.

We analyzed this plasmonic filtering inside the tapered part of the NSOM tip in more detail by simulating the field propagation of the linearly and the azimuthally polarized modes using in the last 2.4 μm of the tapered fiber tip. We separately launched a linearly polarized mode and an azimuthally polarized mode with the same initial power and monitored its power during propagation towards smaller tip diameters. Obviously, both modes experience loss caused by interaction with the metal walls. Initially the power of the linearly polarized mode first drops more rapidly as its normal field components penetrate deeper into the metal (Fig. 5(b)). Respective effective mode indices have a bigger imaginary part for the linearly than for the azimuthally polarized mode (e.g. at $d = 2.4$ μm linear: $\text{Im}(n_{\text{eff,lin}}) = 2.42 \cdot 10^{-3}$, azimuthal: $\text{Im}(n_{\text{eff,azi}}) = 1.47 \cdot 10^{-3}$). Only for smaller diameters $d$ this evolution is inverted. The normalized transmission for azimuthal polarization drops rapidly at an aperture tip diameter of approximately $d = 1.0$ μm, indicating a quasi-cutoff. In contrast the cutoff of the linearly polarized mode happens for much smaller radii [33]. Making use of the numerically determined field structures of the interacting modes we again analyze our experimental results for the most interesting aperture diameter of $d = 1.4$ μm in more detail. To determine the content of azimuthally polarized light present in the experiment we assumed the total measured field to consist only of the linearly and the azimuthally polarized mode. Using these numerically determined modes we tried to reproduce the observed intensity pattern as $I_{\text{tot}}(x,y,z) \propto |E(x,y,z)_{\text{azi,taper}} + \alpha \cdot E(x,y,z)_{\text{lin,taper}}|^2$ and compared this result with the measured data (Fig. 3(b)). The best matching of the simulation with the experiment was obtained for α = 0.9 ± 0.1. Consequently a ratio of about

$$\frac{\int_{-\infty}^\infty dx \int_{-\infty}^\infty dy \, |E(x,y,z)_{\text{azi,taper}}|^2}{\int_{-\infty}^\infty dx \int_{-\infty}^\infty dy \, |E(x,y,z)_{\text{azi,taper}} + \alpha \cdot E(x,y,z)_{\text{lin,taper}}|^2} = \frac{1}{1+|\alpha|^2} = 55\% \pm 5\%$$

of the emitted power belonged to an azimuthally polarized mode, a value, which is consistent with the result of the full FDTD simulation.

The tip aperture diameter of $d = 1.4$ μm, for which we observed the strongest azimuthally polarized field, is only slightly smaller than the wavelength of operation (1550 nm). However, the electric field does not fill the whole tip completely and the full outer FWHM of the complete doughnut-shaped electric field distribution is Ø ≈1 μm (Fig. 6(c)). The strong magnetic field in the center of this doughnut that we find in the FDTD simulations (see Fig. 6(a)) is predominantly polarized in propagation direction along the tip (here x). Its power distribution has a much smaller diameter of FWHM ≈450 nm (Fig. 6(b)). Its longitudinal ($H_x$) component is even tighter focused to FWHM ≈300 nm, which is deeply subwavelength.

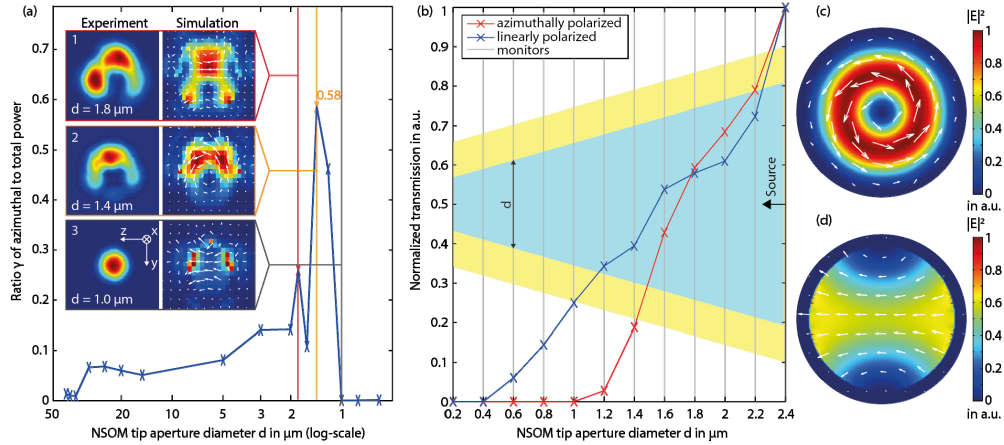

Fig. 5. (a) Ratio γ of azimuthal field power to the total field power derived from the 3D FDTD simulation along the full tip in the uv-plane (see Fig. 4(a)). The maximum portion of azimuthal polarization is achieved for $d = 1.4$ μm. Inset (a1 – a3) Experimental and simulated field distribution $|(\mathbf{E})(x,y,z)|^2$ for the respective tip aperture diameters in the yz-plane (see Figs. 2, 3, and 4(a)). The simulation shows the field distribution at the apex including the emitted and evanescent fields. Especially the simulation in the inset (a3) shows strong near-field enhancement at the edge of the tip apex, which corresponds to a marginal remaining azimuthally polarized part. Within the scope of the measurement precision no azimuthal component was detected for $d = 1.0$ μm. All overlaid arrows show Re($\mathbf{E}$)($y,z$)) indicating the polarization of the fields emitted from the tip. (b) 3D FDTD simulation of the propagation of an azimuthally and a linearly polarized mode through the final section of a truncated cone resembling the plasmonic modal filter at the end of the tip. We separately injected the two modes with equal power at a tip aperture diameter $d = 2.4$ μm. The azimuthally polarized mode first experiences less loss than the linearly polarized one, but then it runs into a quasi-cutoff for $d ≈ 1.0$ μm. The linearly polarized mode has its cutoff only for $d ≈ 0.4$ μm. (c) – (d) Respective source modes for the simulation in (b).

In all measurements (Fig. 3) and simulations of the full system (Fig. 4) where azimuthally polarized fields are detected we constantly observe an asymmetric intensity distribution in y-direction at the tip apex, e.g. for a tip aperture diameter of $d = 1.4$ μm (Fig. 3(b) and Fig. 4(b)). Always more power is concentrated at one side of the tip apex corresponding to the inner side of the bend. Such asymmetry can only be caused by the bend itself, because all other parts are cylindrically symmetric. As only two modes reach the end of the taper and as the respective beating length $L = \lambda/2\, \Delta n ≈ 900$ μm with $\Delta n = $ Re($n_{\text{eff,lin}}$) − Re($n_{\text{eff,azi}}$) = $8.63 \cdot 10^{-4}$ is much larger than the length of the taper (≈360 μm), after the bend itself the initial asymmetry remains conserved.

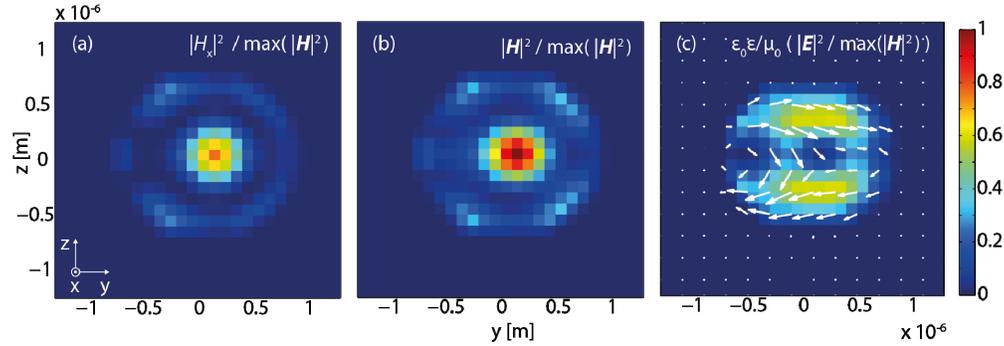

Fig. 6. Comparison of the energy density of the magnetic and the electrical fields for the tip aperture ($d = 1.4$ μm) obtained via 3D FDTD simulation of the bent and tapered NSOM tip (see Fig. 4(a)). All images are normalized to the maximum magnetic energy density ½ $\mu_0$ (max $|\mathbf{H}|^2$). (a) As it is characteristic for an azimuthally polarized electric field distribution, the magnetic field is mainly polarized in propagation direction ($H_x$) and concentrated in a subwavelength spot with a diameter of FWHM ≈300 nm. The total magnetic field (b), has a FWHM ≈450 nm. (c) The electrical energy density is displayed together with arrows representing the real part of its transverse components (y,z), clearly indicating the dominating azimuthal field distribution with an outer diameter of FWHM ≈1 μm.

## 4. Conclusion

In conclusion we have demonstrated the first fiber tip source for azimuthally polarized light ($|\mathbf{E}_{azi}|^2 / |\mathbf{E}_{tot}|^2 \approx 55\% \pm 5\%$) based on a custom processed NSOM tip with a tip aperture diameter $d = 1.4$ μm $< \lambda_0 = 1.55$ μm. To fully understand the generation of this polarization distribution we have analyzed the evolution of the optical field in the tip with 3D FDTD simulations. We found a specific resonant tip aperture diameter that selectively suppresses other supported modes, but transmits the azimuthally polarized field distribution together with a residual portion of linearly polarized light. When launching an optimized input polarization into the single mode optical fiber that is connected to the NSOM tip, the maximum transmission is observed for a field distribution at the tip apex, which is a mixture of an azimuthally and a linearly polarized mode. Based on a full field analysis using FDTD calculations we attribute the generation of these azimuthal components to a three-step process:

1. A linearly polarized fundamental fiber mode with an orientation orthogonal to the symmetry plane of the bent and tapered fiber tip is injected into the main taper region.

2. As the cylindrical symmetry of the optical fiber is broken in the bend of the tip, azimuthal field components are generated by reflection on the concave metallized walls of the bend.

3. The last part of the fiber tip, resembling a truncated, metal coated cone [34,35], acts as a plasmonic mode filter that selectively transmits azimuthal and linear polarization.

This combined process generates an azimuthally polarized field distribution at a tip aperture with a diameter of $d = 1.4$ μm and for an optimized in-coupling polarization. This aperture diameter is indeed only slightly smaller than the wavelength of operation. However, according to literature [36–40], all other fiber based or photonic crystal fiber (PCF) based sources of azimuthally polarized light emit much less confined field distributions. Although in some cases impressive conversion ratios are obtained (e.g. 99.8% [36]) those fiber based sources seem not to be suitable for operation in a NSOM. This is mainly because respective core diameters (e.g. ø ≈8 μm for a doped fiber [36] or ø ≈23 μm for a PCF [38,40]) are rather large, while the limit of size for NSOM applications lies in the size of the electromagnetic field distributions to investigate. In addition a conversion into [36,37,39] or the filtering for azimuthally polarized modes [38,40] has previously only been reported for multi-mode fibers with tens of microns diameter. Hence, a preservation of the azimuthally polarized mode over longer distances or sharp bends, as they occur in NSOM configurations like the one presented might not be feasible.

Up to date the strongest confinement of azimuthally polarized light has been experimentally obtained with high purity by focusing macroscopic azimuthally polarized beams from a large free-space setup with high numeric aperture [22,23]. The width of such generated field distributions is not much bigger than the confinement presented in this article, but still limited by the Abbe diffraction limit. However, free-space propagating fields are not able to couple to the near-field of a sample, hence excluding typical NSOM applications. This intrinsically inhibits the application for exciting or probing of non-radiative near-fields, e.g. in nano- or micro-photonics or plasmonics. In addition numerical investigations have recently revealed that even deep-subwavelength (FWHM $\geq \lambda_0/19$) field structures can be generated in fiber tips provided that they can be decorated with 3d nanostructures [17]. Azimuthally polarized electric field distributions exhibit a strong magnetic field mainly oscillating in propagation direction, respectively along the NSOM tip. This magnetic field is even more localized (FWHM $\approx 450$ nm $\approx \lambda_0/3.5$) compared to the electric field (FWHM $\approx 1$ μm $\approx \lambda_0/1.5$), which will potentially allow for investigating magnetic near-field distributions. Furthermore, our device supports only one additional linearly polarized mode, which is possible to separate for some sample and measurement configurations.

For measurements, a structure under investigation can be either excited selectively with the field emerging from such a tip or its near-field can be collected and measured [5,23,41–44]. Our scheme, based on a standard NSOM tip, is easy to implement, reproduce and apply in comparison to alternative approaches [11,18] and requires less sophisticated fabrication techniques, offering a broad possible range of applications.

## Acknowledgments

The authors thank V. H. Schultheiss, S. Dobmann and A. Regensburger for inspiring discussions paving the way for new ideas. This work was supported by the Cluster of Excellence Engineering of Advanced Materials (EAM), Erlangen. We acknowledge the use of facilities at the Max Planck Institute for the Science of Light, Erlangen. D.P. and A.K. also acknowledge funding from the Erlangen Graduate School in Advanced Optical Technologies (SAOT) by the German Research Foundation (DFG) in the framework of the German excellence initiative and Friedrich-Alexander-University Erlangen-Nuremberg (FAU) within the funding programme Open Access Publishing.